\documentclass[letterpaper, 10 pt, conference]{ieeeconf}  

\IEEEoverridecommandlockouts                              

\makeatletter
\let\NAT@parse\undefined
\makeatother

\usepackage{graphicx}          
\usepackage{algpseudocode}
\usepackage{amsmath,amssymb}
\usepackage{amsfonts}
\usepackage{bm}
\usepackage{algorithm}
\usepackage{algpseudocode}
\usepackage{xcolor}
\usepackage{tikz}
\usepackage{soul}
\usepackage{hyperref}
\usepackage{subfigure}
\usetikzlibrary{shapes,arrows,calc}
\usetikzlibrary {shapes.geometric}
\tikzstyle{block} = [draw, rectangle, 
    minimum height=7em, minimum width=8em, align=center]
\tikzstyle{block2} = [draw, rectangle, 
    minimum height=2em, minimum width=3em, align=center]
\tikzstyle{sum} = [draw, circle, node distance=1cm]
\tikzstyle{sum2} = [draw, scale=0.4, circle, node distance=1cm]
\tikzstyle{input} = [coordinate]
\tikzstyle{output} = [coordinate]
\tikzstyle{pinstyle} = [pin edge={to-,thin,black}]

\newtheorem{rem}{Remark}
\newtheorem{definition}{Definition}
\newcommand{\figurename}[1]{Fig.#1}

\title{\LARGE \bf
Meta-learning of data-driven controllers with automatic model reference tuning: theory and experimental case study
}

\author{Riccardo Busetto$^{1}$, Valentina Breschi$^{2}$, Federica Baracchi$^{1}$ and Simone Formentin$^{1}$
\thanks{$^{1}$R. Busetto, F. Baracchi and {S. Formentin} are with Dip. di Elettronica, Informazione e Bioingegneria, Politecnico di Milano, Milano, Italy.
        {\tt\small simone.formentin@polimi.it}}%
\thanks{$^{2}$V. Breschi is with Dept. of Electrical Engineering, Eindhoven University of Technology, Eindhoven, The Netherlands .
        {\tt\small v.breschi@tue.nl}}%
        \thanks{
		This work has been partially supported by FAIR (Future Artificial Intelligence Research) project, funded by the NextGenerationEU program within the PNRR-PE-AI scheme (M4C2, Investment 1.3, Line on Artificial Intelligence),
		by the Italian Ministry of Enterprises and Made in Italy in the framework of the project 4DDS (4D Drone Swarms) under grant no. F/310097/01-04/X56, and
		by the PRIN PNRR project P2022NB77E “A data-driven cooperative framework for the management of distributed energy and water resources” (CUP: D53D23016100001), funded by the NextGeneration EU program.
		}
}

\begin{document}
\pgfdeclarelayer{background}
\pgfdeclarelayer{foreground}
\pgfsetlayers{background,main,foreground}

\maketitle
\thispagestyle{empty}
\pagestyle{empty}

\begin{abstract}
Data-driven control offers a viable option for control scenarios where constructing a system model is 
expensive or time-consuming. Nonetheless, many of these algorithms are not entirely automated, often necessitating the adjustment of multiple hyperparameters through cumbersome trial-and-error processes and demanding significant amounts of data. In this paper, we explore 
a meta-learning approach to leverage potentially existing prior knowledge about analogous (though not identical) systems, aiming to reduce both the experimental workload and ease the tuning of the available degrees of freedom. We validate this methodology through an experimental case study involving the tuning of proportional, integral (PI) controllers for brushless DC (BLDC) motors with variable loads and architectures.
\end{abstract}

\section{Introduction}
Direct, data-driven control is getting increasing traction in control research and practice (see, e.g., the special issues containing \cite{dorfler2023a,dorfler2023b} and \cite{bazanella2023data}), with the promise of sparing control designers from any explicit modeling step whenever little to no prior is available on the system to be controlled apart from a set of input/output measurement. Examples of these strategies range from the recently proposed data-enabled predictive control approaches (see \cite{Verheijen2023} for a review) to \textquotedblleft classical\textquotedblright techniques, like the \emph{iterative feedback tuning} (IFT) \cite{hjalmarsson2002iterative} method and the \emph{virtual reference feedback tuning} (VRFT) \cite{campi2002virtual} approach, to mention a few. Nonetheless, these techniques often do not leverage all information at one's disposal for practical control design. In particular, the controlled system often belongs to a broader family of systems constructed and calibrated to be nominally the same and, hence \textquotedblleft similar\textquotedblright \ by design (this is, for instance, the case for mass-produced devices).\\ 
While this similarity between data-generating systems and/or tasks is largely employed within the machine-learning community (see \cite{rivolli2022meta,Vanschoren2019} and references therein for an overview of the so-called \emph{meta-learning} approach), this additional information is seldom exploited for control design with few exceptions. The \emph{meta-learning} rationale is leveraged in \cite{richards2022control,arcari2023bayesian,park2022meta,Xin22} toward enhancing models, only eventually employed for control design. Hence, in these works the added information stemming from similarity is still used to model rather than to achieve the final control objective. A shift in this paradigm is already performed in \cite{guo2023imitation}, where the objective is reconstructing rather than designing a controller from data, and then ultimately made in \cite{busetto2023meta}. In this approach, data from an unknown system are combined with controllers learned for other unknown (yet similar) plants toward directly designing a controller from data within the VRFT framework. \\
This technique is at the core of our work and, as the traditional VRFT method, it grounds on recasting the control design problem into a controller identification one, by leveraging a user-defined \emph{reference model} that dictates the behavior one aims to attain in closed-loop. This last element is knowingly critical in model-reference control design, as its choice might lead to disruptive consequences on the closed-loop behavior (up to instability) if wrongly performed \cite{van2011data}. To cope with this limitation, several techniques have been proposed over the years, ranging from the use of reference governors to boost the performance attained when selecting a \textquotedblleft conservative\textquotedblright \ reference model \cite{piga2017direct}, to the alternated technique proposed in \cite{Campestrini2011} to cope with non-minimum phase systems and the general framework introduced in \cite{breschi2023autoddc}.\\      
Leveraging the direct, meta-control design framework proposed in \cite{busetto2023meta}, our contribution fits this scenario from two perspectives. From a methodological standpoint, we enhance the flexibility of the core meta-learning approach proposed in \cite{busetto2023meta} by introducing a procedure for auto-tuning the model reference based on a set of user-defined, \emph{soft constraints} on the desired target behavior. This novel auto-tuning problem is formulated by leveraging the bi-level structure introduced in \cite{breschi2023autoddc} while incorporating information on the \emph{attainability} of a certain closed-loop behavior based on the data available in the meta-dataset. From a technical perspective, we showcase the effectiveness of both the core approach proposed in \cite{busetto2023meta} and the novel auto-tuning strategy introduced in this paper in an experimental case study, toward improving the performance of the baseline proportional-integral (PI) controller available within the considered experimental setup for field-oriented control (FOC).\\
The paper is organized as follows. Section~\ref{sec:problem} introduces the considered setting and the problem we aim to tackle. To address it, we firstly summarize the approach proposed in \cite{busetto2023meta} in Section~\ref{sec:background}, of which we then outline the extension with reference model auto-tuning in Section~\ref{sec:autotuning}. The experimental setup considered to validate our work is described in Section~\ref{sec:experimenta_setup}, followed by a presentation and discussion of the obtained results in Section~\ref{sec:experimental_results}. The paper ends with some concluding remarks and directions for future work. 
\section{Setting \& Goal}\label{sec:problem}
Consider an \emph{linear}, \emph{time-invariant}, \emph{single-input single-output} (SISO) 
system $\mathcal{G}$. Let its input/output dynamic be ideally described by the difference equation\footnote{$q^{-1}$ indicates the back-shift operator, i.e., $q^{-i}u(t)=u(t-i)$.}
\begin{equation}\label{eq:system_dyn}
\mathcal{G}:~~y^{\mathrm{o}}(t)=G(q^{-1})u(t),
	\end{equation}
where $u(t) \in \mathbb{R}$ and $y^{\mathrm{o}}(t) \in \mathbb{R}$ indicate the input and (noise-free) output of the system at time $t \in \mathbb{N}$, respectively. 
\begin{figure}[!tb]
    \centering
    \scalebox{.85}{\begin{tikzpicture}
        \node[coordinate] (reference) {};
        \node[coordinate,right of=reference,node distance=.5cm] (aid1) {};
        \node[draw,circle,right of=aid1,node distance=.5cm] (sum1) {};
        \node[draw,rectangle,node distance=2cm,minimum height=2.5em,minimum width=2.5em,right of=sum1] (controller) {$\mathcal{C}(\alpha)$};
        \node[coordinate,node distance=1cm,right of=controller] (aid2) {};
        \node[draw,rectangle,right of=controller,node distance=1.5cm,minimum height=2.5em,minimum width=2.5em] (plant) {$\mathcal{G}$};
        \node[coordinate,right of=plant,node distance=1cm] (aid3) {};
        \node[coordinate,below of=aid2,node distance=1cm] (aid4) {};
        \node[draw,rectangle,below of=aid4,node distance=1cm,minimum height=2.5em,minimum width=2.5em] (referenceMod) {$\mathcal{M}(\varphi)$};
        \node[draw,circle,right of=aid3,node distance=1.5cm] (sum2) {};
        \node[coordinate,right of=sum2,node distance=1cm] (output) {};
        \draw[->] (reference) -- node[yshift=.2cm,near start] {$r(t)$}(sum1);
        \draw[->] (sum1) -- node[yshift=.2cm]{$e(t)$}(controller);
        \draw[->] (controller) -- (plant);
        \draw[->] (plant) -- node[yshift=.2cm]{$y(t)$}node[near end,yshift=.2cm,xshift=.35cm]{-}(sum2);
        \draw[->] (sum2) -- node[yshift=.2cm,near end,xshift=.2cm]{$\varepsilon(t)$}(output);
        \draw[-] (aid3) |- (aid4);
        \draw[->] (aid4) -| node[xshift=-.2cm,yshift=.7cm]{-}(sum1);
        \draw[->] (aid1) |- (referenceMod);
        \draw[->] (referenceMod) -| node[yshift=.2cm,near start]{$y^{d}(t)$}(sum2);
    \end{tikzpicture}}\vspace{-.2cm}
    \caption{Closed-loop matching scheme, with $y(t)$ being the, possibly noisy, closed-loop output and $e(t)=r(t)-y(t)$ being the corresponding tracking error, and $\varepsilon(t)$ the matching error, for $t\geq 0$.}\label{fig:scheme_nometa}\vspace{-.2cm}
\end{figure}
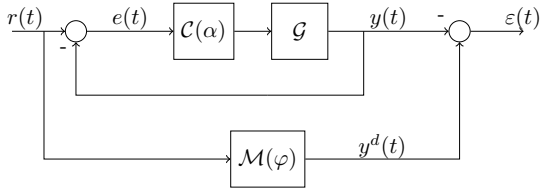
Given a user-defined reference $r(t) \in \mathbb{R}$ and a vector of pre-fixed (rational) basis functions $\beta(q^{-1}) \in \mathbb{R}^{n_{\alpha}}$, our goal is to design a \emph{parametric} controller within the class: 
\begin{equation}\label{eq:controller_class}
    \mathcal{C}(\alpha):~~u(t)=\underbrace{\alpha^{\top}\beta(q^{-1})}_{:=C(q^{-1};\alpha)}(r(t)-y(t)),
\end{equation}
such that the closed-loop system in \figurename{\ref{fig:scheme_nometa}} matches a desired behavior $y^{d}(t)$. The latter is here assumed to be dictated by a \emph{reference model} belonging to the class
\begin{equation}\label{eq:reference_model}
\mathcal{M}(\varphi):~~y^{d}(t)=M(q^{-1};\varphi)r(t),
\end{equation}
with $\varphi \in \mathbb{R}^{n_{\varphi}}$ being a set of user-defined parameters shaping the target closed-loop response.

Let us assume that $G(q^{-1})$ mapping the system's input/output relationship is \emph{unknown}. Nonetheless, suppose that we have access to a (finite) set of input/output measurements $\mathcal{D}_{T}=\{\tilde{u}(t),\tilde{y}(t)\}_{t=1}^{T}$  to accomplish our goal, with\footnote{We denote a white sequence with mean $\mu$ and variance $\sigma^{2}$ as $wn(\mu,\sigma^{2})$.}
\begin{equation}\label{eq:noisy_measurements}
    \tilde{y}(t)=y^{\mathrm{o}}(t)+v(t),~~~v(t)\sim wn(0,\sigma^{2}),
\end{equation}
being the noisy output we can measure from\footnote{Either from open-loop or closed-loop experiments.} $\mathcal{G}$ when we excite it with $\{\tilde{u}(t)\}_{t=1}^{T}$. 

In addition to this dataset, let us assume to have information from $M \geq 1$ systems that are $\varepsilon$-similar to the one we aim at controlling according to the following definition.
\begin{definition}[\cite{busetto2023meta}]\label{def:1}
		Two systems within the class $\mathcal{G}$ in \eqref{eq:system_dyn} with input/output relationships dictated by $G_{1}(q^{-1})$ and $G_{2}(q^{-1})$, respectively, are said to be $\varepsilon$-similar if the following property holds:
		\begin{equation}\label{eq:varepsilon_similarity}
			\Delta G(q^{-1})=\|G_{1}(q^{-1})-G_{2}(q^{-1})\|_{2}\leq \varepsilon,~~\mbox{with }\varepsilon>0.
		\end{equation}
	\end{definition}
\medskip
Specifically, suppose we have access to a dataset $\mathcal{D}_{T}^{i}=\{\tilde{u}_{i}(t),\tilde{y}_{i}(t)\}_{t=1}^{T}$ collected under the same experimental conditions\footnote{Namely, exciting all the systems with the same input sequence when data collection is performed in open-loop, or considering the same references when collecting data in closed-loop.} considered when gathering $\mathcal{D}_{T}$. Moreover, assume that these data have already been used to tune $M$ distinct controllers $\{C_{i}(q^{-1})\}_{i=1}^{M}$ all belonging to \eqref{eq:controller_class}, each tuned for the corresponding closed-loop system to match a \emph{prefixed} reference model in the class \eqref{eq:reference_model}, i.e., $\mathcal{M}(\bar{\varphi})$ with $\bar{\varphi} \in \mathbb{R}^{n_{\varphi}}$ fixed by the designer beforehand. Finally, suppose that these controllers have all been deployed and we have collected the resulting (closed-loop) outputs over $T^{\mathrm{cl}}$ steps, i.e., $\{y_{i}^{\mathrm{cl}}(t)\}_{t=1}^{T^{\mathrm{cl}}}$ for $i=1,\ldots,M$. We hence assume to have access to:
\begin{enumerate}
    \item a set of input/output data $\mathcal{D}_{T}^{i}$, for $i=1,\ldots,M$;
    \item the set of controllers $\{C_{i}(q^{-1})\}_{i=1}^{M}$ designed with these data;
    \item the resulting closed-loop responses $\{y_{i}^{\mathrm{cl}}(t)\}_{t=1}^{T^{\mathrm{cl}}}$ when tracking the reference of interest $\{r(t)\}_{t=1}^{T^{\mathrm{cl}}}$, with $i=1,\ldots,M$. 
\end{enumerate}
These elements constitute what we refer to from now on as the \emph{meta-dataset} $\mathcal{D}^{\mathrm{meta}}$, that we will exploit in combination with $\mathcal{D}_{T}$ (by considering $\mathcal{D}_{T} \cup \mathcal{D}^{\mathrm{meta}}$) to attain two concurrent objectives. On the one hand, we have the goal of $(i)$ tuning a controller in the class \eqref{eq:controller_class} such that the achieved closed-loop response matches the desired one. On the other hand, we wish to $(ii)$ reduce the burden on the user-side, by autotuning the parameters of the reference model \eqref{eq:reference_model} looking at performance improvement and target achieving.
    
\section{Direct meta-control design: an overview}\label{sec:background}
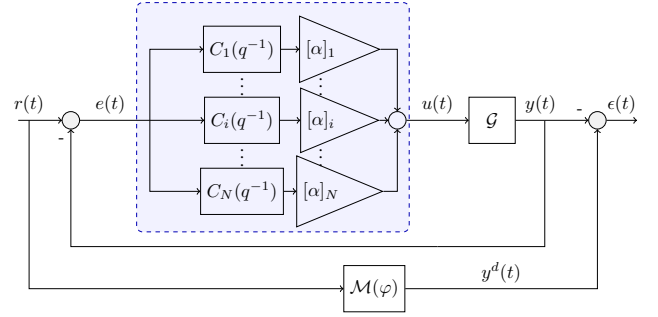
\begin{figure}[!tb]
    \centering
    \scalebox{.7}{
    \tikzstyle{block} = [draw, fill=black!5, rectangle, 
    minimum height=2.5em, minimum width=2.5em]
\tikzstyle{sum} = [draw, fill=black!5, circle, node distance=1cm]
\tikzstyle{input} = [coordinate]
\tikzstyle{output} = [coordinate]
\tikzstyle{pinstyle} = [pin edge={to-,thin,black}]

\begin{tikzpicture}[auto, node distance=2cm]
    \node [input](input){};
    \node [sum, right of=input] (sum1) {};
    \node[coordinate,right of=sum1,node distance=1.5cm] (aid4) {};
    \node[draw,rectangle,minimum height=2.5em, minimum width=2.5em,right of=aid4,node distance=1.75cm] (Ci) {$C_{i}(q^{-1})$};
    \node [draw,isosceles triangle,right of=Ci,node distance=1.5cm, minimum width=3.25em] (alphai) {$[\alpha]_{i}$};
    \node[draw,circle,right of=alphai,node distance=1.45cm] (sum3) {};
    \node[rectangle,above of=Ci,node distance=.75cm] (aid6) {$\vdots$};
    \node[rectangle,above of=alphai,node distance=.75cm] (aid6bis) {$\vdots$};
    \node[rectangle,below of=Ci,node distance=.6cm] (aid7) {$\vdots$};		
    \node[rectangle,below of=alphai,node distance=.6cm] (aid7bis) {$\vdots$};		
    \node[draw,rectangle,minimum height=2.5em, minimum width=2.5em,above of=aid6,node distance=.6cm] (C1) {$C_{1}(q^{-1})$};
    \node[draw,isosceles triangle,right of=C1,node distance=1.5cm, minimum width=3.5em] (alpha1) {$[\alpha]_{1}$};
    \node[draw,rectangle,minimum height=2.5em, minimum width=2.5em,below of=aid7,node distance=.75cm] (CN) {$C_{N}(q^{-1})$};
    \node[draw,isosceles triangle,right of=CN,node distance=1.5cm, minimum width=3.5em] (alphaN) {$[\alpha]_{N}$};
    \node [draw,rectangle,minimum height=2.5em, minimum width=2.5em, right of=alphai, node distance=3.25cm, text width=1em, align=center] (system) {$\mathcal{G}$};
    
    \draw [->] (sum3) -- node[xshift=0cm, name=u, yshift=0cm] {$u(t)$} (system);
    \node [output, right of=system, node distance=1cm] (output) {};
    \node [coordinate, right of=output, node distance=0.5cm] (aid_out) {};
    \node [coordinate, below of=u, node distance=2.7cm] (aid_feedback) {};

    \node[sum, right of=output, node distance=1cm] (sum2) {};
    \node [coordinate, right of=sum2, node distance=.75cm] (aid_out_out) {};
    \node [draw,rectangle,minimum height=2.5em, minimum width=2.5em,below of=aid_feedback, node distance=0.8cm, xshift=-1.2cm](model ref){$\mathcal{M}(\varphi)$};
    \draw [draw,->] (input) -- node[near start, yshift=.cm] (r2) {$r(t)$} (sum1);
    \draw [draw,->] (model ref.east) -|
    node[near start,yshift=0cm] {$y^d(t)$} (sum2);
    \draw [draw,->] (r2) |- (model ref);
    \draw [draw,->] (sum1) -- node [xshift=-0.6cm, yshift=.0cm] {$e(t)$} (Ci);
    \draw [draw,->] (aid4) |- (C1);
    \draw [draw,->] (aid4) |- (CN);
    \draw [draw,-] (system) -- node [xshift=0.2cm] {$y(t)$} (output);
    \draw [draw,-] (output) |- (aid_feedback);
    \draw [draw,->] (output) -- node [pos=.8] {-}(sum2);
    \draw [draw,->] (sum2) -- node [pos=.5] {$\epsilon(t)$} (aid_out_out);
    \draw [->] (aid_feedback) -| node[pos=0.95] {-} (sum1);

    \draw[->] (alpha1) -| (sum3);
    \draw[->] (alphai) -- (sum3);
    \draw[->] (alphaN) -| (sum3);
    \draw[->] (C1) -- (alpha1);
    \draw[->] (Ci) -- (alphai);
    \draw[->] (CN) -- (alphaN);

    \coordinate (aid_meta) at (4.2,2.2);
                
      \begin{pgfonlayer}{background}
        \begin{pgflowlevelscope}{\pgftransformscale{.8}}
            \path (C1.west |- C1.north)+(-0.7,1) node (a) {};
            \path (alpha1.east |- alphaN.east)+(+2.4,-1.3) node (c) {};
            \path (alphaN.south -| alphaN.east)+(+1,-0.3) node (b) {\textcolor{blue!70!black}{$C(\alpha)$}};
            \path[fill=blue!5!white,rounded corners, draw=blue!70!black, dashed]
            (a) rectangle (c);
        \end{pgflowlevelscope}
        \end{pgfonlayer}
\end{tikzpicture}
    }\vspace{-.2cm}
    \caption{Closed-loop matching scheme with the meta-controller. We indicate as $y(t)$ the, possibly noisy, closed-loop output, $e(t)=r(t)-y(t)$ is the corresponding tracking error, and $\varepsilon(t)$ is the matching error, for $t\geq 0$.}
    \label{fig:meta_scheme}\vspace{-.2cm}
\end{figure}
Before introducing the procedure for the concurrent (direct) design of the controller in \eqref{eq:controller_class} and the tuning of the reference model in \eqref{eq:reference_model}, we here summarize the approach proposed in \cite{busetto2023meta} to leverage all information at our disposal for control design. To this end, for now, we neglect the dependence on $\varphi$.

The first step to this end lies in defining the new controller to be designed. By relying on the fact that $\{C_{i}(q^{-1})\}_{i=1}^{M}$ in the meta-dataset belong to \eqref{eq:controller_class}, we structure the new input/output map of the controller as follows\footnote{Given $a \in \mathbb{R}^{n_a}$, $[a]_i$ denotes its $i$-th component, for $i=1,\ldots,n_a$.}:
\begin{equation}\label{eq:meta_controller}
    C(q^{-1};\tilde{\alpha})=\sum_{i=1}^{M} [\tilde{\alpha}]_{i}C_{i}(q^{-1}), 
\end{equation}
which belongs by construction in the class in \eqref{eq:controller_class}. Nonetheless, this alternative structural choice allows us to leverage the assumed $\varepsilon$-similarity of the considered systems to \emph{restrict} the space of admissible parameters $\alpha$. In particular, we impose
\begin{subequations}\label{eq:constriants}
\begin{align}
    &\sum_{i=1}^{M} [\tilde{\alpha}]_{i}=1,\\
    &[\tilde{\alpha}]_{i}\geq 0,~~~i=1,\ldots,M,
\end{align}
\end{subequations}
hence envisioning a controller that is a convex combination of those already available in $\mathcal{D}^{\mathrm{meta}}$ and leading to the scheme in \figurename{\ref{fig:meta_scheme}}. 

Then, we have to define the criterion to be optimized when learning the parameter vector $\tilde{\alpha} \in \mathbb{R}^{M}$. Among others, we consider the same objective function already considered in \cite{busetto2023data}, i.e.,
\begin{equation}\label{eq:VRFT_loss}
    J^{\mathrm{c}}(\tilde{\alpha})\!=\!\!\bigg\|F(q^{-1\!})\!\!\left[M(q^{-1\!};\varphi)\!-\!\frac{C(q^{-1\!};\tilde{\alpha})G(q^{-1\!})}{1\!+\!C(q^{-1\!};\tilde{\alpha})G(q^{-1\!})}\right]\!\!\bigg\|_{2}^{2}\!,
\end{equation}
where $F(q^{-1})$ is a user-defined weighting filter. Nonetheless, we augment it with two regularization terms, respectively accounting for the \emph{performance} of the controllers in the meta-dataset and the similarity of the associated systems to the one we now aim to control. This leads to the following control design problem
    \begin{subequations}\label{eq:ideal_meta_problem}
        \begin{align}
            &\min_{\tilde{\alpha}}~~J(\tilde{\alpha};\mathcal{J},\Delta G)\\
            & ~~ \mbox{s.t. }~C(q^{-1};\tilde{\alpha})=\sum_{i=1}^{M}[\tilde{\alpha}]_{i}C_i(q^{-1}),~~ \sum_{i=1}^{N} [\tilde{\alpha}]_{i}=1,\label{eq:ideal_constr1}\\
            & ~~\qquad~ [\tilde{\alpha}]_{i} \geq 0,~~i=1,\ldots,M,\label{eq:ideal_constr3}
    \end{align}
where 
\begin{equation}\label{eq:ideal_cost}
    J(\tilde{\alpha};\mathcal{J},\Delta G)\!=\!J^{\mathrm{c}}(\tilde{\alpha})+\!
            \lambda_{J}R_{J}(\tilde{\alpha};\mathcal{J})+\lambda_{S}R_{S}(\tilde{\alpha};\Delta G),
\end{equation}
    \end{subequations}
and $\lambda_{J},\lambda_{S}>0$ are two hyper-parameters to be selected by the user, while $R_{J}: \mathbb{R}^{M} \rightarrow \mathbb{R}^{+}$ and $R_{S}: \mathbb{R}^{M} \rightarrow \mathbb{R}^{+}$ are regularization functions of the performance $\mathcal{J}$ and the plants' similarity $\Delta G$. This allows us to embed the theoretical insights provided by \cite[Proposition 2]{busetto2023meta}, while intuitively trying to benefit from information collected from systems in the meta-set similar to the one we aim to control and those controllers that achieve \textquotedblleft good\textquotedblright \ closed-loop performance.
\subsection{Toward practical meta-control design}
The matching loss in \eqref{eq:ideal_cost} still depends on the \textquotedblleft true\textquotedblright \ input/output relationship characterizing the system we aim to control. As the latter is unknown, we exploit the same steps of \cite{campi2002virtual} to translate it into its data-driven counterpart, leading to
\begin{equation}\label{eq:dd_cost}
    J^{\mathrm{c}}(\tilde{\alpha};\mathcal{D}_{T})=\sum_{t=1}^{T}\left[\zeta(t)\left(u^{L}(t)-C(q^{-1};\tilde{\alpha})e_{v}^{L}(t)\right)\right]^{2},
\end{equation}
where $\{u^{L}(t)\}_{t=1}^{T}$ is the set of input data filtered as discussed in \cite{campi2002virtual}, $e_{v}^{L}(t)$ is the filtered counterpart of the \emph{virtual error} $\tilde{e}_{v}(t)=(M(q^{-1})-1)\tilde{y}(t)$ (with $t \in [1,T]$) and $\zeta(t)$ is an \emph{instrumental variable} introduced\footnote{The reader is referred to \cite{Stoica2002} for possible approaches to construct such instrument.} to cope with noise.

While theoretically allowing us to account for the performance of controllers in the meta-dataset and the similarity of the systems they have been designed for to the one we aim to control, the regularization terms are also not yet conceived to be used in practice. Indeed, both $\tilde{J}$ and $\Delta G$ in \eqref{eq:ideal_cost} must be estimated from data. Relying on the intuitions introduced in \cite{busetto2023data}, we here construct their data-driven approximations as:
\begin{align}
    & \mathcal{J}_{i} \approx \tilde{\mathcal{J}}_{i}= \frac{1}{T}\sum_{t=1}^{T}(y^{d}(t)-\tilde{y}_{i}(t))^{2},~~~~i=1,\ldots,M,\\
    & \Delta G_{i} \approx \mathcal{S}_{i}=\frac{1}{T}\sum_{t=1}^{T}(\tilde{y}(t)\!-\!y_{i}^{\mathrm{cl}}(t))^{2},~~i=1,\ldots,M,\label{eq:similarity_data}
\end{align}
hence evaluating the performance of controllers in $\mathcal{D}^{\mathrm{meta}}$ based on the one we have \textquotedblleft experienced\textquotedblright \ when deploying them while assessing similarity via comparing the available (training) outputs $\tilde{y}(t) \in \mathcal{D}_{T}$ with $\tilde{y}_{i}(t) \in \mathcal{D}_{T}^{i}$, for all $t \in [1,T]$ and $i=1,\ldots,M$. This translate into the following direct, \emph{data-driven meta-control} problem:
    \begin{subequations}\label{eq:data_meta_problem}
        \begin{align}
            &\min_{\tilde{\alpha}}~~J(\tilde{\alpha};\tilde{\mathcal{D}}_{T})\\
            & ~~ \mbox{s.t. }~C(q^{-1};\tilde{\alpha})=\sum_{i=1}^{M}[\tilde{\alpha}]_{i}C_i(q^{-1}),~~ \sum_{i=1}^{N} [\tilde{\alpha}]_{i}=1,\label{eq:ideal_constr1}\\
            & ~~\qquad~ [\tilde{\alpha}]_{i} \geq 0,~~i=1,\ldots,M,\label{eq:ideal_constr3}
    \end{align}
where $\tilde{\mathcal{D}}_{T}=\mathcal{D}_{T} \cup \mathcal{D}^{\mathrm{meta}}$ and
\begin{equation}\label{eq:data_cost}
    J(\tilde{\alpha};\tilde{\mathcal{D}}_{T})\!=\!J^{\mathrm{c}}(\tilde{\alpha};\mathcal{D}_{T})+\!
            \lambda_{J}R_{J}(\tilde{\alpha};\tilde{\mathcal{J}})+\lambda_{S}R_{S}(\tilde{\alpha};\mathcal{S}).
\end{equation}
    \end{subequations}
\begin{rem}[Choice of regularization] Among possible choices, we consider the following regularization terms:
\begin{equation}
    R_{J}(\tilde{\alpha};\tilde{\mathcal{J}})=\|\tilde{\alpha}\|_{\tilde{\mathcal{J}}}^{2},~~R_{S}(\tilde{\alpha};\mathcal{S})=\|\mathcal{S}^{\top}\tilde{\alpha}\|_{1},
\end{equation}
toward enforcing sparsity in $\tilde{\alpha}$ based on similarity, while reducing the weights of those controllers that (based on our experience) perform worse than others.
\end{rem}
\section{Meta-AutoDDC}\label{sec:autotuning}
Even allowing meta-information incorporation, the control problem~\eqref{eq:data_meta_problem} still demands the final user to prefix the reference model dictating the design objective. Nonetheless, such a choice is knowingly challenging, as it has to be carried out with little knowledge of the system to be controlled. Performing this \textquotedblleft blind\textquotedblright \ choice can lead to severe consequences (up to closed-loop instability) if the user selects an excessively demanding target behavior, or poorer performance when \textquotedblleft conservative\textquotedblright \ choices are made. 

Instead of asking the user to make such a choice, we here require the designer only to specify a range of \textquotedblleft acceptable\textquotedblright \ closed-loop behaviors, that we use (along with the information in the meta-dataset) to tune the reference model $M(q^{-1};\varphi) \in \mathcal{M}(\varphi)$ along with the controller. Along the line of \cite{breschi2023autoddc}, we translate these \textquotedblleft soft\textquotedblright \ specifications into bounds on  the parameters of the reference model, i.e.,
\begin{equation}
    \varphi \in \Phi \subset \mathbb{R}^{n_{\varphi}},
\end{equation}
and we make the \emph{shift} from a \emph{prefixed} to a \emph{flexible} reference model by defining the following (auto-tuning) bi-level optimization problem:
\begin{subequations}\label{eq:auto_meta_problem}
    \begin{align}
    & \min_{\varphi}~~J^{\mathrm{auto}}(\varphi;\tilde{\mathcal{D}}_{T}^{\mathrm{cal}})\\
    & ~~~\mbox{s.t }~\varphi \in \Phi,\\
    & \qquad ~~ \tilde{\alpha}(\varphi)=\arg\min_{\tilde{\alpha} \in \mathcal{A}} J(\tilde{\alpha};\varphi,\tilde{\mathcal{D}}_{T}),
    \end{align}
\end{subequations}
where $\mathcal{A} \subset \mathbb{R}^{n_{\alpha}}$ encodes the constraints in \eqref{eq:constriants} and $\tilde{\mathcal{D}}_{T}^{\mathrm{cal}\!}\!=\!\mathcal{D}_{T^{\mathrm{cal}}} \cup \mathcal{D}^{\mathrm{meta}\!}$ and $\mathcal{D}_{T^{\mathrm{cal}}}\!=\!\{\tilde{u}^{\mathrm{cal}}(t),\tilde{y}^{\mathrm{cal}}(t)\}_{t=1}^{T^{\mathrm{cal}}}$ is a calibration dataset disjoint from $\mathcal{D}_{T}$ (i.e., $\mathcal{D}_{T^{\mathrm{cal}}} \cap \mathcal{D}_{T\!} =\! \emptyset$). Meanwhile, $J^{\mathrm{auto}}(\varphi;\tilde{\mathcal{D}}_{T})$ maps the performance we aim at embed in the auto-tuned reference model while avoiding the choice of possibly unachievable targets. Differently from the approaches summarized in \cite{breschi2023autoddc}, we have additional information to shape this loss. Indeed, thanks to the meta-information we assume to gather from \emph{similar} systems, we have a valuable prior for shaping the choice of $M(q^{-1};\varphi)$. 

Specifically, we exploit this prior by partitioning  $J^{\mathrm{auto}}(\varphi;\tilde{\mathcal{D}}_{T}^{\mathrm{cal}})$ as follows:
\begin{equation}\label{eq:cost_auto}
    J^{\mathrm{auto}}(\varphi;\tilde{\mathcal{D}}_{T}^{\mathrm{cal}})=J^{\mathrm{perf}}(\varphi;\tilde{\mathcal{D}}_{T}^{\mathrm{cal}})+J^{\mathrm{meta}}(\varphi;\mathcal{D}^{\mathrm{meta}}),
\end{equation}
where (as in \cite{breschi2023autoddc}) the first term is defined as
\begin{equation}
J^{\mathrm{perf}}(\varphi;\mathcal{D}_{T}^{\mathrm{cal}})\!=\!\!\sum_{t=1}^{T^{\mathrm{cal}}} \!Q(\tilde{e}_{v}^{\mathrm{cal}}(t))^{2\!}+R(\Delta u^{\mathrm{cal}}(t;\tilde{\alpha}(\varphi)))^{2\!\!},
\end{equation}
with $Q,R>0$ being hyper-parameters that the user can tune to steer the choice of the reference model, while
\begin{subequations}
\begin{equation}
    \Delta u^{\mathrm{cal}}(t;\tilde{\alpha}(\varphi))=u^{\mathrm{cal}}(t;\tilde{\alpha}(\varphi))-u^{\mathrm{cal}}(t-1;\tilde{\alpha}(\varphi)),
\end{equation}
is the difference between inputs reconstructed with $C(q^{-1};\tilde{\alpha}(\varphi))$ at two consecutive time instants by using the calibration dataset (with $u^{\mathrm{cal}}(0;\tilde{\alpha}(\varphi))=0$) and
\begin{equation}
    \tilde{e}_{v}^{\mathrm{cal}}(t)=\left[M(q^{-1};\varphi)-1\right]\tilde{y}^{\mathrm{cal}}(t),
\end{equation}
for $t=1,\ldots,T^{\mathrm{cal}}$.
\end{subequations}

The second penalty $J^{\mathrm{meta}}(\varphi;\tilde{\mathcal{D}}_{T}^{\mathrm{cal}})$ in the calibration loss is instead chosen as
\begin{equation}
    J^{\mathrm{meta}}(\varphi;\mathcal{D}^{\mathrm{meta}})=\sum_{i=1}^{M}\sum_{t=1}^{T^{\mathrm{cl}}} \mathcal{S}_{i}(M(q^{-1};\varphi)r(t)-y_{i}^{\mathrm{cl}}(t))^{2},
\end{equation}
where $\mathcal{S}_{i}$ is defined as in \eqref{eq:similarity_data} and $r(t)$ is the user-defined reference to be tracked. This choice embeds the intuition that the more similar a system used to construct the meta-dataset is to the one we aim to control, the more the closed-loop response of the former encodes a behavior that we can attain with the latter. According to this intuition, based on similarity we can obtain a data-based prior on responses that might be achievable by controlling the system of interest, which we hence exploit in the definition of our calibration loss.
\section{BLDC motor meta-control: experimental setup}\label{sec:experimenta_setup}
\begin{figure}[!tb]
	\centering
	\includegraphics[width=.8\linewidth, trim=0cm 10cm 0cm 0cm,clip]{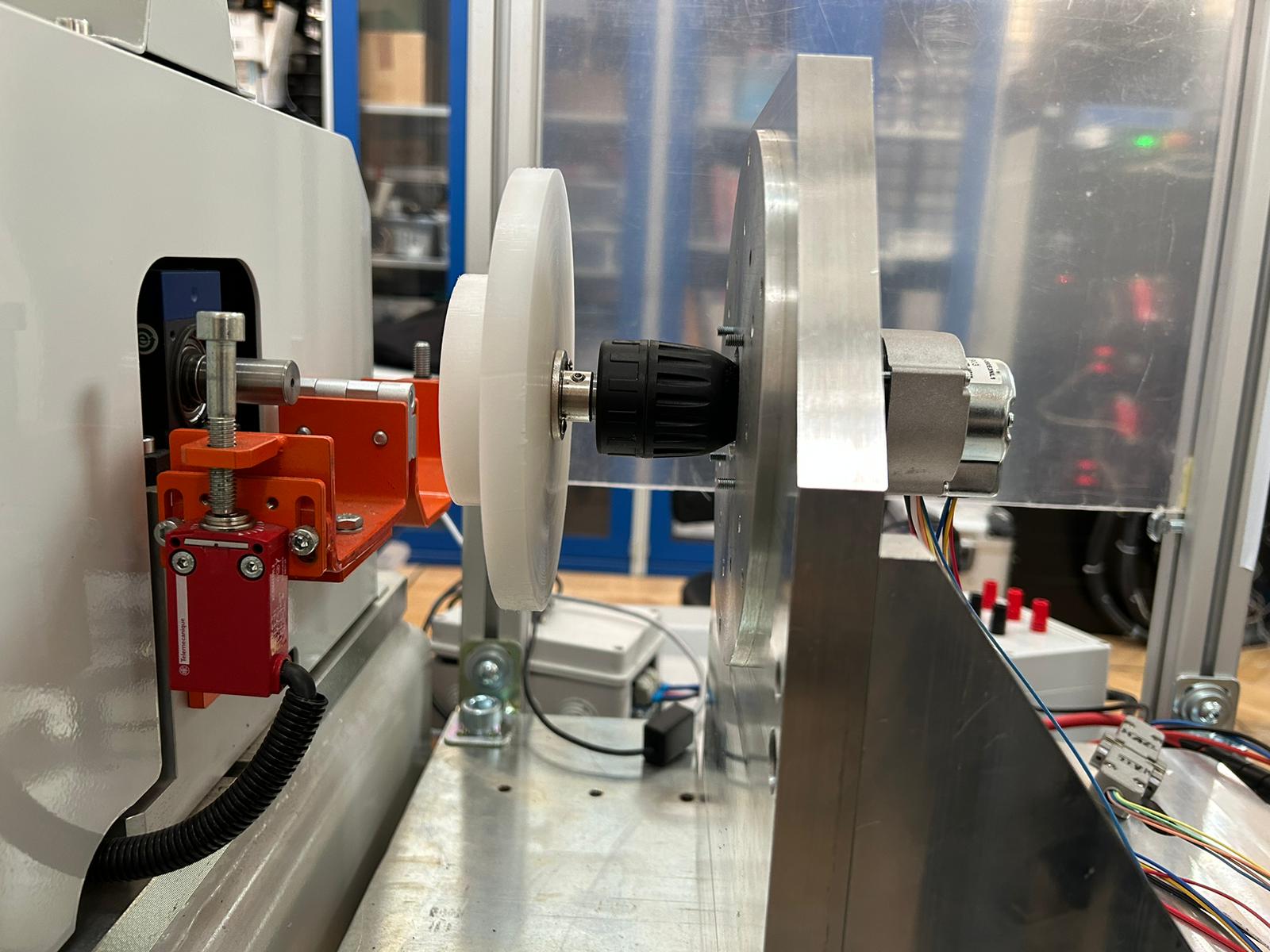}\vspace{-.2cm}
	\caption{Experimental setup for a specific configuration. Note that, the motor and the inertial load are connected through a spindle.}
	\label{fig:setup}\vspace{-.2cm}
\end{figure}
\addtolength{\tabcolsep}{-1.5pt}  
\begin{table*}[!tb]
   \caption{Considered motor+load configurations. When two values are reported, two load discs have been combined.}
   \label{tab:loads}\vspace{-.2cm}
   \centering 
   \begin{tabular}{lcccccccccc}
   \hline
   & \textbf{1} & \textbf{2} & \textbf{3} & \textbf{4} & \textbf{5} & \textbf{6} & \textbf{7} & \textbf{8} & \textbf{9} & \textbf{10}\\
   \hline
   \textbf{r} [m] & $0.024$, $0.029$ & $0.0447$ & $0.0447$ & $0.0543$ & $0.0447$ , $0.0447$ & $0.069$ & $0.0342$ , $0.069$ & $0.0741$ & $0.0741$ , $0.0447$  & $0.0841$\\ 
   \textbf{m} [kg] & $0.055$, $0.073$ & $0.115$ & $0.225$ & $0.174$ & $0.115$ , $0.225$ & $0.207$ & $0.088$ , $0.207$ & $0.212$ & $0.212$ , $0.115$ & $0.212$\\
   \textbf{I} $\cdot$ 10$^{-3}$ [kg $\cdot$ m$^2$] & $0.0465$ & $0.1149$ & $0.2248$ & $0.2565$ & $0.3397$ & $0.4928$ & $0.5442$ & $0.5820$ & $0.6969$ & $0.7497$\\
   \hline
   \end{tabular}
\end{table*}
\addtolength{\tabcolsep}{1.5pt}  
To assess the practical validity of the proposed approach 
we consider the problem of designing a speed controller within a \emph{Field Oriented Control} (FOC) scheme for \textit{Brushless DC} (BLDC) motors. Our goal is to steer the velocity $y(t)=\omega(t)$~[rpm] of the motor connected to an inertial load to a desired target speed through the quadrature current $u(t)=i_q(t)$~[A]. Our experimental setup, shown in \figurename{\ref{fig:setup}} for a specific configuration, comprises the following.
\begin{itemize}
    \item \emph{Two BLDC motors}, a Shinano Kenshi LA052 (with 4-pole-pairs, rated power of 80 W, rated speed of 3000 rpm, rated voltage 24 V, rated current 7 A) and a Maxon EC-i 40 (with 7-pole-pairs, rated power of 100 W, rated speed of 3920 rpm, rated voltage 18 V, rated current 5.46 A). Both motors are equipped with hall-effect sensors for speed measurement and are rigidly mounted on a vertical clamping. 
    \item \textit{The power supply}, here being a bidirectional DC accounting for regenerative load, with rated power up to 6kW. 
    \item \textit{A three-phase motor controller} available on the demonstration board EVSPIN32G4 by ST Microelectronics for BL motors. The firmware includes the FOC architecture, which has been modified to allow external communication and to enable changes in the controller gains.
    \item \textit{An external computing unit}, namely x64-based PC, Intel(R) Core i7 (2.30 GHz, 8 Cores), 16 GB RAM. 
    \item \textit{A} UART-based ($f_s=1$~[kHz]) \emph{communication network} for exchanging data between the internal controller and the external PC.
\end{itemize}
Based on the FOC structure already available for the system, we focus on designing a \emph{proportional}, \emph{integral} (PI) speed controller, i.e., $C(q^{-1};\alpha) \in \mathcal{C}(\alpha)$ such that
\begin{equation}\label{eq:basis_PI}
\beta(q^{-1})=\begin{bmatrix}
    1\\
    \frac{T_{s}}{2}\frac{1+q^{-1}}{1-q^{-1}}
\end{bmatrix},
\end{equation}
where $T_s=1$~[ms] is the sampling time of the BLDC motor. The solution of the meta-control design problem in \eqref{eq:data_meta_problem} has been obtained via the CVX package \cite{gb08,cvx} on an Intel(R) Core(TM) i7-10875H CPU @ 2.30GHz processor with 16 GB of RAM running MATLAB R2023b, while the auto-tuning procedure is carried out on the same machine via SMGO-$\Delta$. 

To build the meta-dataset, the two motors have been employed in combination with up to $10$ different inertial loads (their description is available in Table~\ref{tab:loads}), leading to a total of $20$ possible configurations. Among them, $M=10$ configurations are randomly selected to construct the meta-dataset $\mathcal{D}^{\mathrm{meta}}$, while the $10$ remaining ones are used to assess the effectiveness of the meta Auto-DDC approach proposed in Section~\ref{sec:autotuning}. The latter compose what we refer to in the following as the test set\footnote{For each testing configuration, the corresponding element in this set represents $\mathcal{D}_{T}$.} $\mathcal{D}^{\mathrm{test}}$. 
\subsection{Data collection}
\begin{figure}[!tb]
    \centering
    \includegraphics[scale=.6,trim=19.5cm 35cm 20.5cm 35cm,clip]{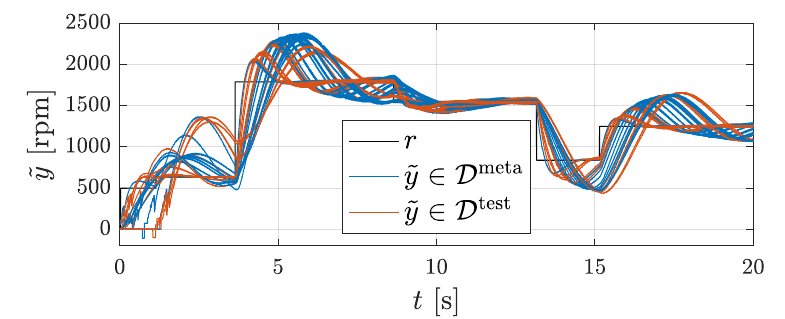}\vspace{-.2cm}
    \caption{Datasets: responses comprised in the meta dataset $\mathcal{D}^{\mathrm{meta}}$ (used to compute \eqref{eq:similarity_data}) and the training sets $\mathcal{D}^{\mathrm{test}}$.}
    \label{fig:responses_fixed_controller}\vspace{-.2cm}
\end{figure}

For all $20$ configurations data are collected in closed-loop by considering a sequence of steps as the reference signal, using a conservative PI controller with proportional and integral gains equal to $6.836 \cdot 10^{-1}$ and $1.892 \cdot 10^{-3}$, respectively. This choice allows us to carry out \textquotedblleft safe\textquotedblright \ data collection campaigns (i.e., without run-away conditions), which is instead difficult if open-loop experiments are performed. The data collection campaign was performed over $20$~[s], with data gathered with a sampling frequency of $1$ [kHz], resulting in datasets of length $T=20000$. The measured outputs resulting from this data-collection campaign are shown in \figurename{\ref{fig:responses_fixed_controller}}, highlighting the significant difference between the considered configurations (and, hence, our exploration capabilities). In particular, those trajectories that highlight a slower response of the underlying system are those associated with configurations with higher inertia connected to the motor.
\begin{rem}[On the use of closed-loop experiments]
    While the theoretical work in \cite{busetto2023meta} postulates the use of data collected in open-loop, this is not possible here for safety reasons. Nonetheless, this choice does not undermine our procedure, as noise in direct control design is handled via an instrumental variable approach (see \eqref{eq:dd_cost}).  
\end{rem}
\subsection{Construction of the meta-dataset}\label{sec:meta_constr}
From the $20$ available configurations (see Table~\ref{tab:loads}), $M=10$ randomly selected motors have been used to construct the meta-dataset $\mathcal{D}^{\mathrm{meta}}$. For each of these motors, we have thus tuned a PI controller in \eqref{eq:controller_class} by exploiting the technique proposed in \cite{busetto2023data}. In particular, the parameters of the controller have been tuned by solving
\begin{equation*}
    \min_{\alpha_{i}} \sum_{t=1}^{T^{\mathrm{cl}}} (y_{i}^{\mathrm{cl}}(t;\alpha_{i})-y^{d}(t))^{2},   
\end{equation*}
via $n_{\mathrm{itr}}=30$ iterations\footnote{The search is restricted to $[10^{-2},30]$ and $[10^{-4},10^{-1}]$ for the proportional and integral gain, respectively.} of SMGO-$\Delta$ \cite{sabug2022smgo} over a closed-loop experiment of the duration of $5$~[s] (i.e., $T^{\mathrm{cl}}=5000$), with $y_{i}^{\mathrm{cl}}(t;\alpha_{i})$ and $\bar{y}^{d}(t)$ being the attained and desired closed-loop responses to a step speed reference $r$ of amplitude $1500$ [rpm] and
\begin{equation}\label{eq:reference_model_fixed}
y^{d}(t)=\underbrace{\frac{0.609q^{-1}}{1-0.9391q^{-1}}}_{=M(q^{-1})}r(t),
\end{equation}
for all $i=1,\ldots,M$. Along with the optimal controller parameters $\{\alpha_{i}^{\star}\}_{i=1}^{M}$, for the construction of the meta-dataset we have further stored the associated closed-loop responses, i.e., $\{y_{i}^{\mathrm{cl}}(t;\alpha_{i}^{\star})\}_{t=1}^{T^{\mathrm{cl}}}$ for $i=1,\ldots,M$.
\section{Experimental results}\label{sec:experimental_results}
\begin{figure}[!tb]
        \centering
        \begin{tabular}{c}
            \subfigure[SMGO]{\includegraphics[scale=1,trim=0cm 0cm 44cm 72.5cm,clip]{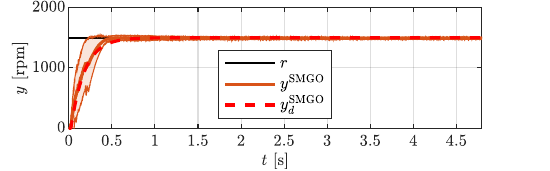}}\vspace{-.2cm}\\
            \subfigure[META\label{fig:meta_notuning}]{\includegraphics[scale=1,trim=0cm 0cm 44cm 72.5cm,clip]{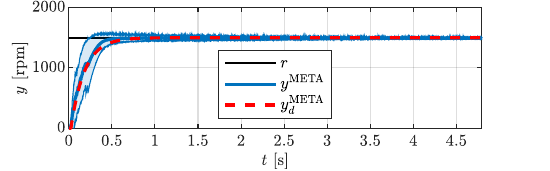}}\vspace{-.2cm}\\
            \subfigure[VRFT]{\includegraphics[scale=1,trim=0cm 0cm 44cm 72.25cm,clip]{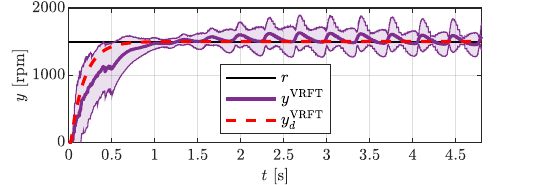}}
        \end{tabular}\vspace{-.2cm}
            \caption{Comparison of direct, data-driven techniques: set point (black line) and desired response (dashed red line) \emph{vs} mean (colored line) and standard deviation (shaded area) of the closed-loop responses attained with different controllers across the test motors.}\label{fig:meta-results}
\end{figure}
We initially focus on testing the effectiveness of the meta-control design technique proposed in \cite{busetto2023meta} within our experimental setting, by fixing the reference model as in \eqref{eq:reference_model_fixed}. In particular, we consider the same step reference used for the construction of the meta-dataset (see Section~\ref{sec:meta_constr}), comparing the performance attained by solving \eqref{eq:data_meta_problem} with those achieved by designing the PI controller with the standard VRFT approach \cite{campi2002virtual} and the method proposed in \cite{busetto2023data} (already used to tune the controllers in the meta-dataset). Note that, when solving \eqref{eq:data_meta_problem} we fix $\lambda_{J}=10^{16}$ and $\lambda_{S}=10^{17}$ since the first term in the loss can attain values to orders up to $10^{20}$
. The results in \figurename{\ref{fig:meta-results}} clearly show that both the meta-design approach proposed here and the technique introduced in \cite{busetto2023data} outperform the classical VRFT approach, as further confirmed by the indicator reported in Table~\ref{tab:index_comparison1}. While performance is slightly better when using SMGO-$\Delta$, this result comes at the price of a larger variability in performance across the testing configurations. The latter is slightly smaller when using the proposed meta-control strategy, spotlighting its capability to lead to more consistent closed-loop performance when the controlled system changes while belonging to the considered class. At the same time, as reported in Table~\ref{tab:index_comparison1}, the slight performance improvement achieved with the approach presented in \cite{busetto2023meta} comes at the price of a more demanding experimental campaign. Indeed, the calibration of each new controller with SMGO-$\Delta$ (as by \cite{busetto2023data}) requires a $5$~[s] long \emph{closed-loop, online experiment} for each of the $n_{\mathrm{itr}}=30$ SMGO-$\Delta$ iterations. On the other hand, provided the meta-dataset, the proposed meta-control strategy requires only the \emph{offline} collection of two input/output datasets (one of which is used to construct the instrument used in \eqref{eq:dd_cost}), here gathered with experiments of $20$~[s].  
\begin{table}[!tb]
    \caption{Comparison of direct, data-driven techniques: performance indexes and data-collection time per BLDC motor.}
    \label{tab:index_comparison1}
    \centering
    \begin{tabular}{lccc}
    \hline
     &  SMGO-$\Delta$ & META & VRFT\\
     \hline
     $\|y^{d} - y\|_2$~[rpm] & 
     8\textbf{8.3} $\pm$ 32.0
     & 90.5 $\pm$ \textbf{29.7} &
     193.4 $\pm$ $116.2$\\
     $\|r - y\|_2$ [rpm] & 
     \textbf{221.2} $\pm$ 60.1
     & 223.0 $\pm$ \textbf{58.5} &
     333.4 $\pm$ 97.4\\
     $\|\Delta u\|_2$ [A] & 0.11 $\pm$ \textbf{0.02}
     & \textbf{0.10} $\pm$ \textbf{0.02} &
     0.13 $\pm$ \textbf{0.02}\\
     Collection time [s] & 150~[s] & \textbf{40}~[s] & \textbf{40}~[s]\\
     \hline
     \end{tabular}
\end{table}
\subsection{On the performance of meta Auto-DDC}
\begin{figure}[!tb]
    \includegraphics[scale=1,trim=0cm 0cm 44cm 72cm,clip]{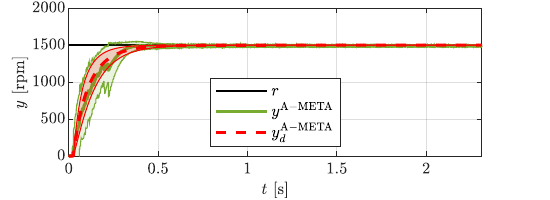}\vspace{-.2cm}
    \caption{Performance of the meta-controller with auto-tuning: mean desired response (dashed red line) and actual closed-loop output (green line) with their standard deviation (shaded areas) across the test motors.}\label{fig:auto1}\vspace{-.2cm}
\end{figure}
\begin{figure*}[!tb]
        \centering
        \begin{tabular}{ccc}
            \subfigure[Mismatching error: $\|y^{d}-y\|_{2}$]{\includegraphics[scale=.9,trim=0cm 0cm 48.9cm 71cm,clip]{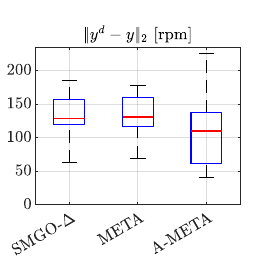}}
            &
            \subfigure[Tracking error: $\|r-y\|_{2}$]{\includegraphics[scale=.9,trim=0cm 0cm 48.9cm 71cm,clip]{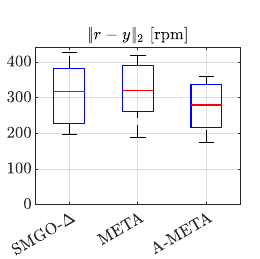}}
            &
            \subfigure[Input effort: $\|\Delta u\|_{2}$]{\includegraphics[scale=.9,trim=0cm 0cm 48.9cm 71cm,clip]{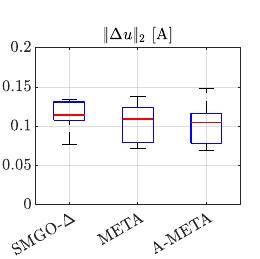}}
            \end{tabular}\vspace{-.2cm}
            \caption{Comparison of non-flexible direct, data-driven techniques and the proposed meta-autoDDC approach: performance indicators across the test motors over 3~[s] long experiments.}\label{fig:usage_vsperformance_similarity}
\end{figure*}
We now show how performance changes (and is enhanced) when using the meta-design approach with reference model auto-tuning (denoted as A-META), considering a first-order reference model parameterized as
\begin{equation}
    M(q^{-1};\varphi)=\frac{(1-\varphi)q^{-1}}{1 - \varphi q^{-1}},
\end{equation}
and allowing $\varphi \in \Phi \equiv [0.9550,0.9991]$, corresponding to enforcing a desired settling time within the interval $ [0.1, 5]$ [s].
To search the optimal reference model according to \eqref{eq:cost_auto} within this exploration space, we solve \eqref{eq:auto_meta_problem} via $n_{\mathrm{itr}}=30$ of SMGO-$\Delta$, fixing $Q=10^{8}$ and $R=10^{3}$. \figurename{\ref{fig:auto1}} showcases the closed-loop performance of the meta-autoDDC controller against the desired model reference responses resulting from the auto-tuning procedure. This result suggests improved matching performance on average, especially at the end of the transient, along with a slight reduction in the closed-loop settling time, which drops from approximately $1$~[s] without auto-tuning (see \figurename{\ref{fig:meta_notuning}}) to around 0.5~[s].

Given the results in \figurename{\ref{fig:meta-results}} and Table~\ref{tab:index_comparison1}, we further compare the outcome of the meta-autoDDC approach introduced in Section~\ref{sec:meta_constr} with those of the methods proposed in \cite{busetto2023data,busetto2023meta}. Given the settling time evidenced by \figurename{\ref{fig:meta-results}} and \figurename{\ref{fig:auto1}}, without loss of generality this analysis is restricted to closed-loop experiments of $3$~[s].

\figurename{~\ref{fig:usage_vsperformance_similarity}} shows the statistics of the performance indicators already introduced in Table~\ref{tab:index_comparison1} over the considered test set of motors. Enhancing the flexibility of the design strategy leads to consistent improvements in the mean of both the tracking and mismatching errors, along with the input effort. The only counter-effect of increasing the flexibility of the reference model is represented by the slight increase in variability of the associated box plots for the mismatching error and the input effort. Nonetheless, the first result is mainly due to the closed-loop behavior of a single BLDC motor, whose response is slightly slower than the one dictated by the auto-tuned reference model. Instead, the second is due to two motors requiring slightly more control effort at the gain of a more adherent tracking of the reference behavior.

\section{Conclusions}
In this work, we have extended a recently proposed meta-design approach toward enhancing closed-loop performance. This result is achieved by increasing the reference model's flexibility, while exploiting all components and, thus, information within our meta-dataset. Apart from these methodological contributions, a key aspect of this work lies in the experimental validation of all meta-design direct approaches considered throughout the paper. This comparison shows that using a more flexible meta-control scheme can bring advantages in performance while reducing the burden on the user side.  

From a methodological standpoint, future work will be devoted to extending the auto-tuned, meta-control design rationale to other direct, data-driven control techniques. From a more practical perspective, future work will focus on easing the tuning of regularization parameters and testing the proposed design rationale on other real-world problems.


\bibliographystyle{IEEEtran}
\bibliography{main}

\end{document}